\documentclass[structabstract]{aa}  
\usepackage{graphicx}
\usepackage{psfig}
\usepackage{txfonts}
\usepackage{natbib}
\begin{document}
   \title{Investigating the driving mechanisms of coronal mass ejections}

   \author{Chia-Hsien Lin
          \and
           Peter T. Gallagher
          \and
           Claire L. Raftery
          }

   \institute{Astrophysics Research Group,
              School of Physics, Trinity College Dublin,
              Dublin 2, Ireland }

   \date{}

 
  \abstract
   {} 
   {The objective of this study was to examine 
the kinematics of coronal mass ejections (CMEs) using 
EUV and coronagraph images,
and to make a quantitative comparison with a number of theoretical models.
One particular aim was to investigate the acceleration profile of CMEs
in the low corona.
    }
   {We selected two CME events for this study,
    which occurred on
    2006 December 17 (CME06) and 2007 December 31 (CME07).
    CME06 was observed using 
    the EIT and LASCO instruments on-board SOHO, while
    CME07 was observed using the SECCHI imaging suite on STEREO.
    The first step of the analysis was 
    to track the motion of each CME front
    and derive its velocity and acceleration.
    We then compared the observational kinematics,
    along with the information of the associated X-ray emissions 
    from GOES and RHESSI,
    with the kinematics proposed by three CME models
    (catastrophe, breakout and toroidal instability).
    }
   {We found that CME06 lasted over eight hours while
    CME07 released its energy in less than three hours.
    After the eruption,
    both CMEs were briefly slowed down before being accelerated again.
    The peak accelerations during the re-acceleration phase
    coincided with the peak soft X-ray emissions for both CMEs.
    Their values were $\sim$60~m~s$^{-2}$ for CME06 and 
    $\sim$600~m~s$^{-2}$ for CME07.
    CME07 reached a maximum speed of over 1000~km~s$^{-1}$
    before being slowed down to propagate away at a constant, final speed of 
    $\sim$700~km~s$^{-1}$.
    CME06 did not reach a constant speed
    but was moving at a small acceleration by the end of the observation.
    Our comparison with the theories suggested that
    CME06 can be best described by a hybrid of the catastrophe model and
    breakout model while
    the characteristics of CME07 
    were most consistent with the breakout model.
    Based on the catastrophe model,
    we deduced that the reconnection rate in the current sheet for CME06
    was intermediate,
    the onset of its eruption occurred at a height of $\sim$200~Mm,
    and the Alfv\'en speed and the magnetic field strength
    at this height were approximately 130--250~km~s$^{-1}$ and 7~Gauss, 
    respectively.}
   {}

   \keywords{Sun: coronal mass ejections (CMEs) - 
             Sun: flares - 
             Sun: magnetic fields -
             Sun: corona -
             Sun: atmosphere}

   \maketitle
%

\section{Introduction}

Coronal mass ejections (CMEs) 
are the ejections of
large amount of mass and magnetic flux 
from the Sun to interplanetary space.
The energy released during the process is of the order of
$10^{32} - 10^{33}$ erg.
Statistical studies have reported variations
in the observed properties and projected kinematics of CMEs
\citep[e.g.,][]{CB2004A&A,Zhang_etal2004ApJ}
Nevertheless,
CMEs often rise with a small initial speed of
a few kilometers per second,
followed by a rapid upward expansion 
reaching several hundred or several thousand kilometers per second,
and, eventually, propagate through the interplanetary space at
the ambient solar-wind speed.

Several models have been proposed to
explain the driving mechanism and observed properties of the CMEs.
A valid model must be able to produce the observed kinematics,
dynamics and properties of CMEs.
In Sec.~\ref{sec:model}, 
we give a review of three representative models:
breakout model (BO) \citep{Breakout1999ApJ},
catastrophe model (CA)
\citep[see, e.g.,][]{VanTend_Kuperus1978SoPh,FI1991ApJ,FP1995ApJ,Lin_etal1998ApJ},
and toroidal instability model (TI)
\citep[see, e.g.,][]{Chen_1989ApJ,KT2006PhRvL},
which have been investigated in this study.
Although the profile of the CME kinematics predicted by each model
is qualitatively consistent with the observed three-phase profile
(i.e., slow rise, eruption and slow down),
the exact profile of each model is different.
Several simulations also showed that
a same model can produce different kinematic profiles
by simply varying the initial conditions and magnetic environment
\citep{Lynch_etal2008ApJ,Schrijver_etal2008ApJ,Chen_1989ApJ}.

In addition to the discrepancies among theoretical predictions,
observational studies that were not based on any model,
also reported different functional forms that
best-fitted their respective data.
For instance,
\citet{Sheeley_etal1999}
fitted their data
by
$r(t)=r_0 + 2 r_0 \ln \cosh \left[ v_a(t + \Delta t_0)/2 r_a \right]$;
\citet{Gallagher_etal2003ApJ} formulated
a double exponential function
$a(t)=\left[a^{-1}_r \exp(-t/\tau_r) + 
           a^{-1}_d \exp(t/\tau_d)\right]^{-1}$
to describe the fast rising and decaying acceleration profile;
and \citet{AMN_2002GeoRL} demonstrated that 
their fast CME was best fitted by
the polynomial, $h(t)= h_0 + v_0t + c t^m$, with $m \approx 3.7$,
which is consistent with
the results by \citet{Schrijver_etal2008ApJ}.
Although these studies were able to derive how the velocity and acceleration
evolve through the duration of CMEs,
they did not provide a physical explanation and driving mechanism 
behind such kinematic profiles.
Besides,
while most of these studies chose polynomials or exponential functions or
a simple combination of the two,
there is no guarantee that the chosen function is the only one
that can fit the observational profile.

Studies to verify a specific model by
comparing observations with  models
have also been carried out.
For example, 
by selecting only those that can be characterized as flux-rope type CMEs,
\citet{Krall_etal2001ApJ}
showed that the kinematic curves predicted by the TI model matched
the kinematics in the higher corona but not the lower coronal region.
An extensive examination of a fast CME by \citet{MK2003ApJ}
based on the data from various instruments
showed supportive evidence for the BO model.
However, \citet{Bong_etal2006ApJ} reported that
while the magnetic-field configuration and
X-ray loop field connectivity in their data indicated a breakout process,
the acceleration profile they obtained contradicted the BO model.
The studies to verify the CA model
\citep[e.g.,][]{AMN_2002GeoRL,Schrijver_etal2008ApJ},
however, often deduced their conclusions by
comparing the early-stage expression of CA model
with the acceleration phase and/or the late stage of 
the observed CME kinematic curves.
As already been demonstrated in all observational studies, 
the kinematics of CMEs can change greatly through the duration of the event.
Hence, it is unsurprising that the functional form for the early stage
does not match with the profiles of the later stages.

The objectives of this paper are first
to examine the features and kinematics of CME events,
and then to determine the possible mechanism behind the observed phenomena by
comparing the observations with various theories. 
The theory most consistent with the observation
can thus provide information
on the magnetic environment where the eruption occurs.
We have chosen two CME events for this study.
We first derived their kinematic profiles 
and examined 
the associated X-ray emissions and flares,
and then compared the results with 
those derived from the three different models.
The kinematic profiles from the models have been mainly obtained from 
simulations.
The only stage of which an analytical expression of the kinematics is available 
is the early stage of the eruption.
Hence,
we first focused on
a qualitative comparison of the characteristics between
the observations and different models,
and then carried out an additional quantitative examination
by fitting the observational data with a number of model expressions.
The objectives of the quantitative examination were
to verify the assumptions employed in the models to derive these expressions,
and/or  to obtain an empirical expression for the observed kinematics.
The values of the fitting parameters may be used to infer
the magnetic environment and the initial conditions 
at the onset of the eruption.

The rest of the paper is organized as follows:
the three models
are reviewed
in Sec.~\ref{sec:model}.
The two selected CME events 
and how their motions were traced
are described 
in Sec.~\ref{sec:observation}.
The qualitative and quantitative comparisons with the models are explained
and discussed in
Secs.~\ref{sec:exam_qual} and \ref{sec:exam_quant}.
We conclude our study in Sec.~\ref{sec:conclusion}.

\section{Review of CME models} \label{sec:model}
%
\subsection{Catastrophe model (CA)} \label{sec:mdl_CA}
In a two-dimensional CA model \citep[e.g,][]{FI1991ApJ,PF_2002A&ARv},
the flux rope is represented by a ring circling the Sun,
and, hence, does not have anchoring ends.
This flux-rope magnetic-field configuration is generated
by placing two line sources with opposite polarities 
on the surface around the Sun.
The magnetic field line connecting the two sources
is initially balanced by a downward force from 
the tension of the field line
and a upward force due to the magnetic pressure gradient.
If the photospheric motion moves the two sources closer,
the tension force would increase,
which pulls the field lines downward.
However,
the downward movement, compressing the field lines, 
subsequently increases the pressure force. 
When the sources move closer than a critical distance,
the pressure force takes over 
and 
the flux rope is propelled and accelerated upward,
resulting in the eruption.
The model shows that,
at the critical point,
the distance of the sources ($d$) and the height of the flux rope ($h$) is
related by $h = d/2 \equiv \lambda_0$.
After the rope rises to a certain height ($h \approx d \approx 2 \lambda_0$),
a current sheet begins to form below it,
which acts as a drag force against the rising of the rope.
Using their 2-D model implemented with an isothermal atmosphere,
\citet{LinForbes2000JGR} showed that
the kinematics of CMEs
is dependent on
the reconnection rate in the current sheet.
In their study,
the reconnection rate is constant along the current sheet,
and is
prescribed by the inflow Alfv\'en Mach number, $M_A$, 
at the midpoint of the current sheet.
%
They found two critical values of $M_{\rm A}$, 0.005 and 0.041,
dividing CMEs into three classes:
oscillatory, with and without a deceleration phase.
Since oscillatory CMEs have never been observed,
the first critical value essentially provides 
a lower limit of $M_{\rm A}$ for a CME to happen.
If the reconnection rate is between the two critical values,
the CME in their simulation
first undergoes a fast acceleration phase for $\sim$12~min after 
the loss of mechanical equilibrium,
followed by a deceleration period ($t=20-100$~min)
and then is re-accelerated again (cf. Fig.~6 in the paper). 
For CMEs with a high reconnection rate ($M_{\rm A} > 0.041$),
the deceleration period would not occur.
To examine the temporal behaviour of the reconnection rate and
the rate of magnetic energy release ($dW_{\rm mag}/dt$),
\citet{PFbook2000} used the electric field at the X point, $E_0$, 
to represent the reconnection rate,
and allowed it to freely evolve.
We can see from their result (Fig.~11.6 in \citet{PFbook2000}) that
the profile of $E_0$ closely follows the profile of $dM_{\rm mag}/dt$
and that the peaks of the two profiles almost coincide.
In other words, we may infer from their result that
the peak of the reconnection rate occur at a similar time as
the peak of $dW_{\rm mag}/dt$.
Based on the above results and
assuming that the rate of thermal-energy change 
($dW_{\rm th}/dt$) follows
the rate of magnetic-energy release \citep{ReevesForbes2005ApJ}
and that $dW_{\rm th}/dt$ can be represented by
the soft X-ray (SXR) emission light curve,
we can deduce that the SXR may reach its emission peak
when the reconnection rate is the highest.
%
We can also deduce from the CA theory that
the SXR and hard X-ray (HXR) emissions, 
which are indications of reconnections and flares,
should begin to increase/occur later than the launching of the CME
(i.e., after the reconnections start in the current sheet).

An analytical expression for the kinematics of a thin flux rope 
(i.e., the radius of the flux rope $a \rightarrow 0$)
{\em before} the formation of the current sheet (i.e, $h/\lambda_0 \leq 2$)
can be derived \citep[e.g.,][]{LinForbes2000JGR,PFbook2000}:
\begin{equation}
\dot{h} \approx \sqrt{\frac{8}{\pi}} V_{\rm A0}
\left[
\ln \left(\frac{h}{\lambda_0}\right) + \frac{\pi}{2} - 
2 \tan^{-1}\left(\frac{h}{\lambda_0}\right)
\right]^{1/2}
+ \dot{h_0},
\label{eqn:CM}
\end{equation}
where $h$ is the height of the flux rope, 
and $h_0$ and $\dot{h_0}$ correspond to 
the initial height and speed, respectively.
$\lambda_0$ is the critical height (i.e., the height at the critical point),
and $V_{\rm A0}$ is the Alfv\'en speed at $h = \lambda_0$.
Starting with Eq.~\ref{eqn:CM},
\citet{PFbook2000} further derived simplified expressions for the ``early'' and ``late'' part
of this stage.
By considering the ``early'' part as before the time reaches the Alfv\'en time scale,
i.e., $t \ll \lambda_0/V_{A0}$,
they obtain:
\begin{equation}
h \simeq \lambda_0 + \dot{h_0} t + 
\frac{4 V_{\rm A0}}{5\sqrt{3\pi}} 
\left(\frac{\dot{h_0}}{\lambda_0}\right)^{3/2} t^{5/2}.
\label{eqn:CM_early}
\end{equation}
The expression for the ``late'' part was derived by assuming
$h/\lambda_0 \gg 1$ but $|\ln h|$ still less than $|\ln a|$:
\begin{equation}
\dot{h} \approx \sqrt{\frac{8}{\pi}} V_{\rm A0}
\left[
\ln \left(\frac{h}{\lambda_0}\right) - \frac{\pi}{2}
\right]^{1/2}.
\label{eqn:CM_late}
\end{equation}

After the formation of the current sheet,
the system becomes too complicated for an analytical expression
of the CME kinematics to be derived.

\subsection{Toroidal instability model (TI)} \label{sec:mdl_TI}
The TI model \citep{Chen_1989ApJ,KT2006PhRvL}
considers a pre-existing coronal flux rope
that is initially in a stable equilibrium,
and aims to explain what initiates the eruption of the loop.
There are two dominant forces acting on the flux rope.
One is an outward Lorentz self-force (hoop force) resulting from 
the current flowing in the flux rope,
and the other is a compressing Lorentz force due to 
the interaction between 
the poloidal magnetic fields outside the flux rope and
the current inside the flux rope.
If the external magnetic field ($B_{\rm ex}$) 
decreases sufficiently fast 
from the flux rope and vanishes in the infinity,
the compressing Lorentz force would 
not be strong enough to counter-act a sudden increase of the hoop force
resulting from, for instance, a surge of the current in the flux rope.
The flux rope would then rapidly and continuously expand, 
leading to a CME eruption.
The condition of $B_{\rm ex}$ for the instability to happen was
derived by \citet{Bateman1978book}:
$-R d\ln B_{\rm ex}/dR > 3/2$, where $R$ is the major radius of the flux rope.
We can infer from the above description that
TI theory does not require reconnections to accelerate a CME,
which is a major distinction between TI and the other two models.
Instead, TI proposes that 
a flux rope can erupt and become CME as long as 
the external magnetic fields 
satisfy the instability condition.

The equation of motion derived from the TI model is too complicated to
be solved analytically \citep[see, e.g.,][]{KT2006PhRvL}.
However, 
by considering only two counter-acting Lorentz forces on the rope,
ignoring temporal changes in the external field and 
its resulting flux in the rope,
and assuming a simple profile 
for the external field ($B_{\rm ex}(H)=\hat{B} H^{-n}$),
\citet{KT2006PhRvL} showed that the expression for the beginning stage
of the instability can be approximated as a hyperbolic function:
\begin{equation}
h(\tau) = 
\frac{P_0}{P_1} \sinh(P_1 \tau),\;\;\;\;\;\;\;\;\;\; h\equiv H/H_0 -1 \ll 1,  
\label{eqn:TI_early}
\end{equation}
where $H$ and $H_0$ are the height of the rope and 
the rope height at the on-set of instability;
$\tau$ is the time normalized by Alfv\'en time;
$P_0$ is a parameter composed of
velocity, inductance, rope height and rope radius at the on-set of instability;
and $P_1\equiv \sqrt{n - n_{\rm cr}}$ is associated with 
the profile of the external magnetic field.
\citet{KT2006PhRvL} also presented the profiles 
from their numerical simulations based on the 
three aforementioned simplifications
(i.e., only two acting forces, a constant external field, 
and a simple profile for the external field).
One distinctive feature in their simulation results
is that the acceleration
all shows a fast rise and a more gradual decay.
However,
\citet{Schrijver_etal2008ApJ} demonstrated that
the height--time profile of a TI model can change from 
a hyperbolic function to polynomials
by simply tuning the initial conditions,
which would change the acceleration profile from a fast initial rise to
a more gradual one.

\subsection{Breakout model (BO)} \label{sec:mdl_BO}
One distinctive requirement for the BO model \citep{Breakout1999ApJ} is that
the flux rope is formed in a multi-polar arcade system,
which consists of a low-lying central arcade,
two low-lying side arcades (one on each side of the central arcade),
and a large arcade overlying the three arcades.
The central arcade is what would later become the CME flux rope.
The shearing of the central arcade boundaries causes the arcade to rise,
which subsequently results in the reconnections with the overlying arcade.
As this reconnection (the breakout reconnection)
removes some of the overlying magnetic field lines,
the rising of central arcade becomes faster,
and, in turn, triggers more and faster breakout reconnections.
This process manifests as the eruption phase in a BO scenario.
As the flux rope rises,
it draws the magnetic field lines of the central arcade
close to each other to form a current sheet underneath.
When the current sheet reaches a certain critical dimension,
tearing instability would cause a second set of reconnections, 
which not only release and further accelerate the CME rope,
but also result in post-CME flares.
After this set of reconnections cut off the current sheet,
the magnetic field lines of the side arcades would move in to
form another current sheet,
in which a third set of reconnections would happen
to reform and restore the magnetic fields after the eruption and  
also trigger another series of flaring activities,
which they identified as the ribbon flares.

Since
the breakout reconnection is what allows the flux rope to rise and erupt
in a BO picture,
we expect that, ideally, significant SXR and HXR emissions 
should be detected
from the very beginning
of the loop rising.
If the second set of reconnections,
responsible for cutting off the tethering force from the current sheet
below the CME,
can be equated to the reconnection process in the CA model,
the temporal behaviour of the reconnection rate during this phase
may follow the profile of the corresponding SXR emissions,
as we discussed in Sec.~\ref{sec:mdl_CA}.
Because of the reconnections with both the overlying field lines
and the trailing current sheets,
we can expect indications of flaring activities 
to be detected both below and on the sides of
the CME loop during different stages of the event.
However, we must keep in mind that the emissions and flares during some stages
might not be detectable due to observational limitations.

Although the BO model does not have an analytical expression 
for the kinematics either,
there are three possible profiles produced by
three different simulations
\citep{Lynch_etal2004ApJ,Lynch_etal2008ApJ,DVA2008ApJ}.
While \citet{Lynch_etal2004ApJ} showed that
their 2.5 dimensional simulation result can be fitted by
a single, constant-acceleration profile:
\begin{equation}
h(t) = h_0 + v_0(t-t_0) + \frac{1}{2} a_0 (t-t_0)^2,
\label{eqn:BO}
\end{equation}
the three-dimensional simulation by \citet{Lynch_etal2008ApJ}
produced a more complex kinematic profile
with three phases of constant acceleration:
a long interval of low, constant acceleration
during the foot-point shearing and breakout reconnection,
followed by a short interval of high, constant acceleration
during the second set of reconnections 
(i.e., reconnections in the current sheet formed by 
the central-arcade magnetic field lines),
and finished with a long interval of constant velocity
which simply means zero acceleration.
\citet{DVA2008ApJ} produced 
a slightly different three-phase acceleration profile from their simulations:
a long interval of slow acceleration during the shearing of the footpoints,
a short interval of fast acceleration 
through the breakout and flare reconnections,
and a short interval of fast deceleration at the end indicating
the restoration and reformation of the magnetic field configuration.
%

\section{Observations and data analysis}\label{sec:observation}

The two CMEs selected for this study were observed on
2006 December 17 (CME06) and 2007 December 31 (CME07).
Both events were initiated at the solar limb,
which allow their radial motion to be easily observed.

We have utilized observations from 
EIT (Extreme ultraviolet Imaging Telescope;
\citet{EIT1995SoPh}) $195$\AA {}
wavelength band and
LASCO (Large Angle and Spectrometric Coronagraph;
\citet{LASCO1995SoPh}) for CME06,
and STEREO B
(Solar TErrestrial RElation Observatory; \citet{STEREO2008SSRv})
instruments (EUVI $171$\AA {} wavelength band, cor1 and cor2)
for CME07.
The X-ray emission data were obtained from
GOES
(Geostationary Operational Environmental Satellite;
\citet{GOES1994SoPh})
and RHESSI (Reuven Ramaty High Energy Solar Spectroscopic Imager;
\citet{RHESSI2002SoPh}).

We first applied a temporal running difference
to the time-series images to enhance the contrast,
and then determined the height vs. time relation of the CME 
by tracing the motion of the CME front.
Our determined CME fronts of the two events are
plotted as crosses in Fig.~\ref{fig:img_cme061217} for CME06
and Fig.~\ref{fig:img_euvi} and \ref{fig:img_cme071231} for CME07.
Since the accuracy of the height determination is the most important factor for
the accuracy of the rest of the analysis,
we repeated the height determination at least ten times,
and calculated the standard deviation of the multiple trials.
The standard deviations were of the order of $10^3$km for EIT and EUVI,
and $10^4 - 10^5$km for LASCO and STEREO cor1 and cor2.
We then used the standard deviation plus
the spatial resolution of the images as the uncertainties.
The observational velocity and acceleration were subsequently calculated
by taking the time derivatives of the height.
The errors in the derived velocity and acceleration
were propagated from the height uncertainties
using the IDL function {\tt derivsig()}.
The kinematic profiles of the two CMEs,
along with the error bars and the GOES SXR profiles,
are plotted in Fig.~\ref{fig:kin_cme061217} (CME06) and 
Fig.~\ref{fig:kin_cme071231} (CME07).
In the left panels of both figures,
we plotted only the results of the lower corona
to allow better visibility of the kinematics in this region.

\subsection{2006 December 17 CME (CME06)} \label{sec:obs_061217}

   \begin{figure}
   \centering 
   \includegraphics[width=8cm]{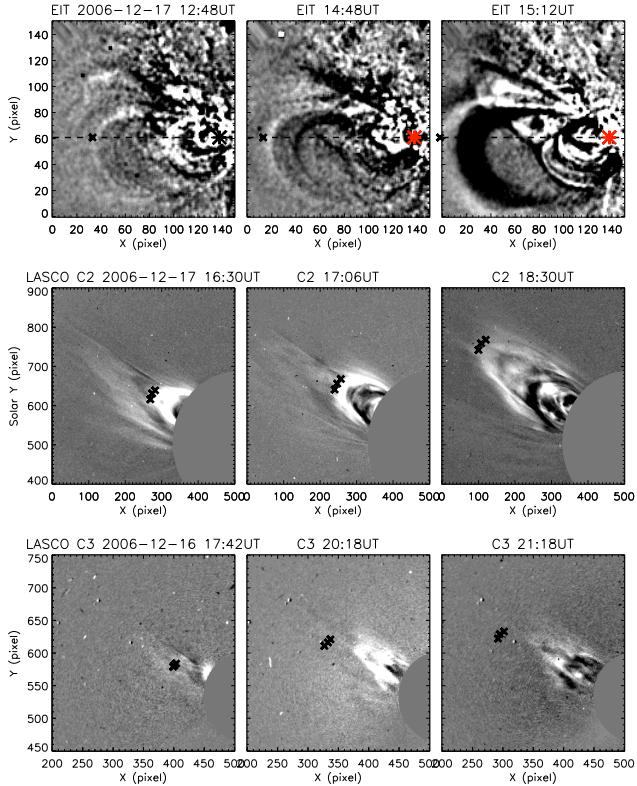}
   \caption{The image sequence of CME06 
            observed by EIT and LASCO C2 and C3.
            The EIT images have been rotated so that the loop top 
            moves along X axis, as indicated by the dash line.
            The pivot of the rotation is plotted as a star,
            which is located at X$\sim$140 along the dash line.
            The crosses in all images mark our determined CME front.
            } 
              \label{fig:img_cme061217}%
    \end{figure}

%

This CME was highly inclined in the low corona
within the field of view of EIT.
Hence, we rotated the EIT images with respect to 
the ``base point''.
The base point
is
the center between
the loop footpoints at the limb.
The line from the loop top to the base point is called ``base line''.
This rotation resulted in the apparent propagation of the loop top
along the X-axis.
Three examples of the rotated EIT running-difference images
are shown in the top row of Fig.~\ref{fig:img_cme061217}.
The base line is plotted as a dashed line,
and the base point is marked as a star.
The images of LASCO are not rotated because
the CME appears to move in the radial direction at such height.

The CME, initially observed by EIT, was embedded among other loops
until it began to rise shortly after 12:00 UT.
The CME was within the field of view (FOV) of EIT until $\sim$15:12~UT.
It was later observed by LASCO C2 between 15:30 and 19:00~UT and
by C3 between 17:00 and 21:18~UT,
after which it became too faint and diffuse to be reliably identified.

\begin{figure}
   \centering 
   \includegraphics[width=8cm]{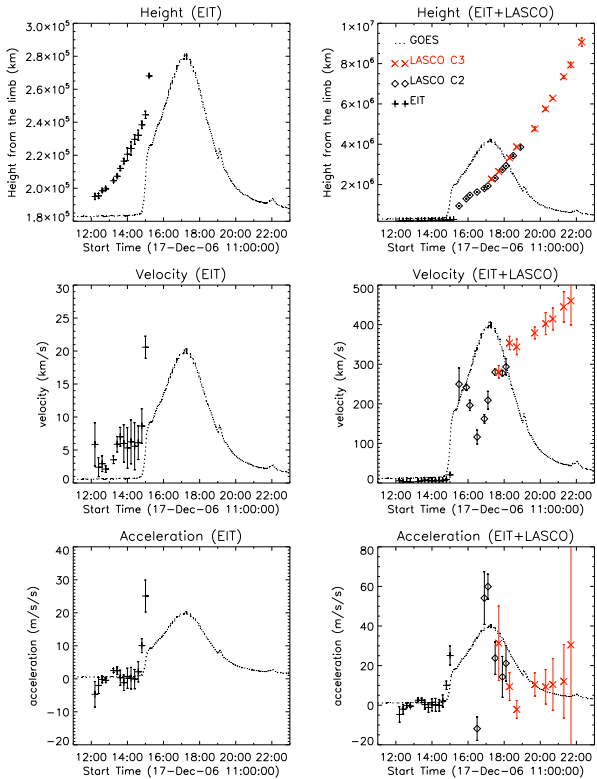}
   \caption{The kinematic profiles of CME06.
            Different symbols represent the observational results from 
            different instruments,
            as denoted in the upper right panel.
            The GOES X-ray emission profile 
            has been scaled to enhance the viewing,
            and hence the exact magnitude is irrelevant.
            The right column shows the combined profile of 
            all three instruments. 
            The left column shows only the results from the low corona 
            to give a better view of the kinematics in this region.}
              \label{fig:kin_cme061217}%
    \end{figure}

The derived kinematics of the CME (Fig.~\ref{fig:kin_cme061217})
shows that
the rise of the CME was initially at a constant speed of a few km~s$^{-1}$
until $\sim$14:30~UT,
when the CME suddenly became eruptive,
indicated by the steep rise of velocity and acceleration.
The eruption was followed by a quick deceleration, 
reaching $\sim -10$~m~s$^{-2}$.
Although the exact duration of the eruptive phase is uncertain
due to the gap between the EIT and LASCO C2 FOV,
the acceleration profile in Fig.~\ref{fig:kin_cme061217} indicates that
the CME turned from eruption to deceleration in $\sim$1.5~hr. 
After the deceleration,
the CME was accelerated again,
and then slowed down to propagate away at a small constant acceleration.

The GOES SXR emissions exhibited two rising phases,
an initial steep rise followed by a more gradual rise.
The first rising phase began when
the CME almost reached the boundary of the EIT FOV ($\ge$14:50~UT).
The second rising phase,
as can be seen in Fig.~\ref{fig:kin_cme061217},
is where the first CME acceleration peak occurred.
The SXR
reached its peak at $\sim$17:15~UT,
when the CME reached the re-acceleration peak
and a flaring loop was detected by EIT.
We also examined the relevant Hinode XRT images,
which showed that this flaring loop was located directly below the erupted loop
and that it contracted slightly before rising to form a cusp-shape loop top.
By comparing the RHESSI and the EIT images,
we found that 
the sources of the RHESSI emission 
(3--6 keV and 6--12 keV) were located
at the top of this post-eruption flaring loop.
The GOES SXR emission finally decreased to a background magnitude 
after 21:00 UT.

\subsection{2007 December 31 CME (CME07)} \label{sec:obs_071231}

   \begin{figure}
   \centering 
   \includegraphics[width=8cm]{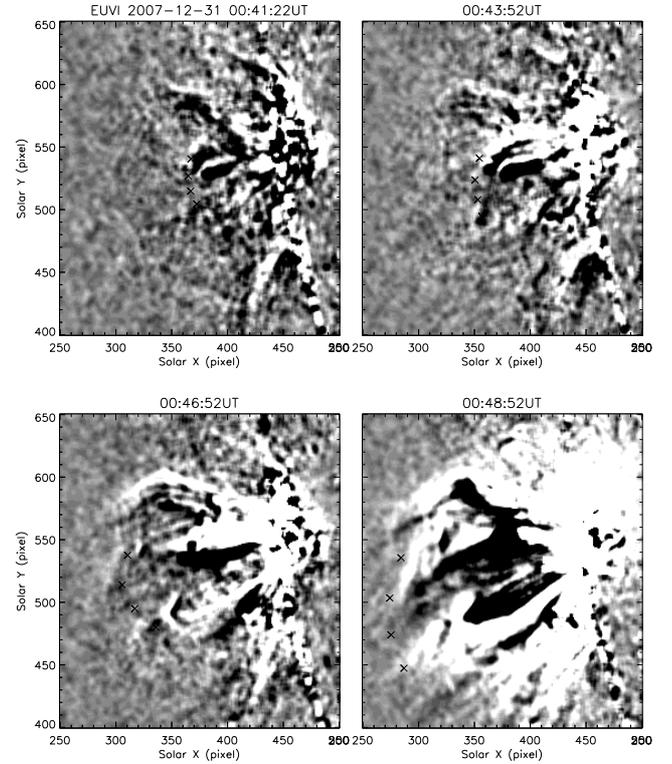}
   \caption{The image sequence to show the formation stage of
            CME07 as observed by STEREO B EUVI 171\AA. 
            The crosses mark our determined CME front.}
              \label{fig:img_euvi}%
    \end{figure}


   \begin{figure}
   \centering 
   \includegraphics[width=8cm]{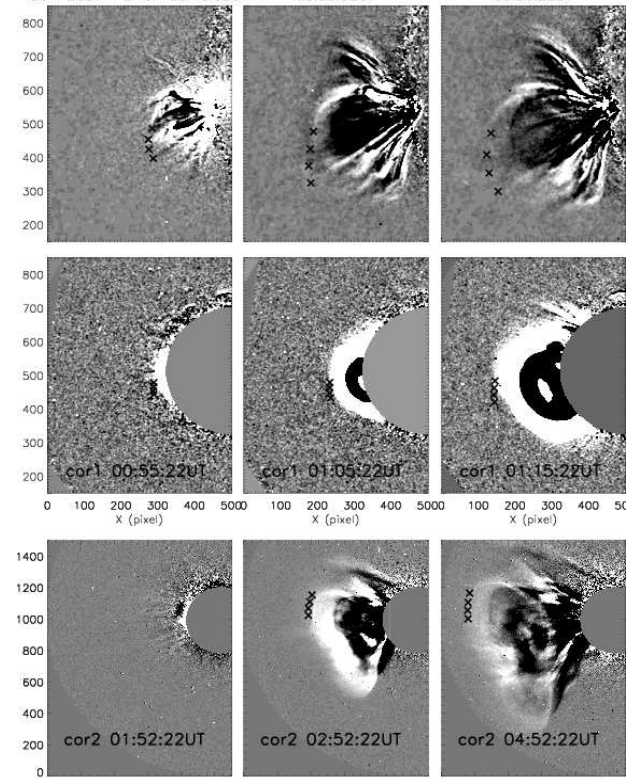}
   \caption{The image sequence of CME07
            observed by STEREO B instruments, EUVI 171\AA, cor1 and cor2.
            The crosses in all images mark our determined CME front.}
              \label{fig:img_cme071231}%
    \end{figure}


Fig.~\ref{fig:img_euvi} shows
the early stages of the CME emergence as observed by EUVI.
The figure revealed that
a system of loop arcades appeared at $\sim$00:41 UT,
and quickly rose radially until 00:48 UT,
after which the top of the arcades became visibly connected 
to form the front of a balloon-shaped CME,
with a round top and a narrow lower part connected to the solar surface.
EUVI also captured some flaring activities and foot-point brightenings
a few minutes before the emergence of the CME.
After the emergence,
Fig.~\ref{fig:img_cme071231} shows that
the CME began to quickly expand sideways 
while the radial rise became slower than previously.
The CME retained the round-shaped top throughout the EUVI and cor1 FOV,
but changed to a shape resembling conjoined balloons
by the time it entered the cor2 FOV.
This suggests that
the flux rope may have been turned, by solar rotation and/or footpoint shearing,
to also show part of the farther side of the rope.

\begin{figure}
   \centering 
   \includegraphics[width=8cm]{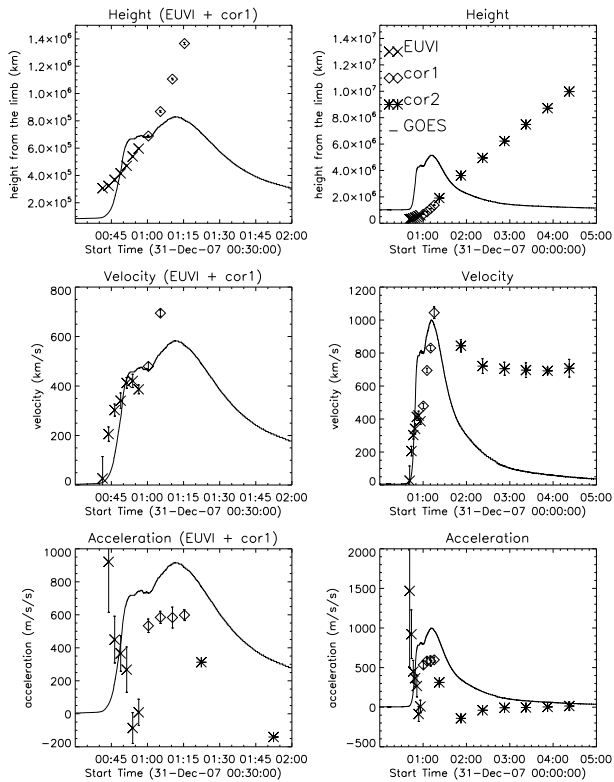}
   \caption{The kinematic profiles of CME07.
            See the caption of Fig.~\ref{fig:kin_cme061217}
             for the description of the figure.}
              \label{fig:kin_cme071231}
    \end{figure}

The velocity profile of CME07 (second row in Fig.~\ref{fig:kin_cme071231}) 
shows that
the first reliable velocity point of the CME is $\sim$200~km~s$^{-1}$.
Unlike CME06, the velocity of the CME07 did not show a clear drop,
but rose monotonically to a peak value of over 1000~km~s$^{-1}$
in under 30 min.
After reaching the peak,
the velocity dropped to a final constant value of $\sim$700~km~s$^{-1}$.
Despite the monotonic rise of the velocity,
the derived acceleration (bottom row of Fig.~\ref{fig:kin_cme071231})
exhibited a decrease during the first 15 min,
and then rose to a plateau value of $\sim$600~m~s$^{-2}$,
which lasted for $\sim$15 min.
After that,
the acceleration dropped to $\sim -150$~m~s$^{-2}$ before
returning to almost zero at the end.
We infer from the high initial velocity
and the decrease of the acceleration at the beginning
that the observations did not detect the initial eruption phase,
during which the CME may have reached an acceleration peak.

The SXR emission from GOES began to rise shortly after 00:40 UT,
reached a peak around 01:10 UT, 
at which time the CME reached its peak velocity,
and dropped to background magnitude roughly after 03:00 UT
when the acceleration of the CME reduced to almost zero.
The analysis of the same event by \citet{Raftery_etal2010apj},
which utilized data from MESSENGER SAX and RHESSI,
showed that the flare/SXR emissions 
actually began much earlier at 00:27~UT 
and lasted until 03:50~UT.
This indicates that the GOES observation was occulted
during the beginning stage of this event.
A study of the RHESSI HXR data of this event by
\citet{Krucker2009SPD,Krucker_etal2009ApJ}
revealed that 
the RHESSI HXR sources were located above 
the SXR sources and were detected at 00:48~UT and 01:02~UT.
The authors deduced that 
the HXR sources may indicate the location of particle acceleration.
Comparing the timing of these two HXR detections 
with the kinematics profile of this CME,
we found that they occurred respectively
at the time of the loop emergence
and at the beginning of the re-acceleration.

\subsection{Qualitative examination}\label{sec:exam_qual}

The first step of this examination was to identify
the properties 
that can be deduced/derived from and 
can be used to distinguish different models.
These properties must also be derivable and/or detectable 
from the observation data we have.
After these distinctive properties were identified,
we then examined the consistency between the features from our observations
and the properties of each models.

A CME can be qualitatively identified as CA type
if it shows the following features:
the CME loop begins to rise  
before an increase in the X-ray emission is observed;
CME-associated flares occur only after the launch of the CME;
and the kinematics resemble any profile produced by the CA model
\citep[e.g.,][]{LinForbes2000JGR}.
If there is no flaring activities around the CME launch site and/or
if the acceleration profile of a CME 
resembles any of the TI simulation results,
it is likely to be a TI type.
However,
we would need a quantitative examination (i.e., fitting) 
to verify whether the profiles proposed by TI
indeed match our observational results.
To classify a CME as a BO type,
we look for the following signs:
significant X-ray emissions and flaring activities
are detected from the very beginning of a CME event;
there are magnetic arcades at the vicinity where the loop is formed;
there are flare ribbons after the CME takes off;
and/or the kinematic profile resembles any of the simulation results.

\subsection{Quantitative Examination}\label{sec:exam_quant}
The mathematical profiles are often derived by imposing some approximations
and assumptions.
Changing the initial conditions and assumptions,
or using different approximations,
can often lead to a different theoretical profile
without changing the model itself \citep[e.g.,][]{Schrijver_etal2008ApJ}.
Besides,
the analytical expressions derivable from the theories are generally limited
to the early stage of a CME process.
An expression for the entire process of CME has not been available due to
the complexity of the realistic models.
Hence, 
the purpose of this quantitative examination is to verify
the mathematical profiles derived from different CME models
and,
when the theoretical expression is unavailable,
to empirically find a
functional form for the kinematics of the CME.
Our strategy was to
first apply several model profiles to fit the height--time data,
then calculate the theoretical velocity and acceleration
by taking the derivatives of the fitting results,
and lastly compare the theoretical and observational kinematic profiles.
%

%
We implemented four fitting models for this exercise,
CA, CAlate\_e, TI and Poly fits.
CA fit was of the form of Eq.~\ref{eqn:CM},
and the fitting parameters were
the critical height $\lambda_0$, initial speed $v_0$, 
and Alfv\'en speed at the critical height $V_{\rm A0}$.
The profile for CAlate\_e fit was Eq.~\ref{eqn:CM_late} 
with a variable exponent, that is,
\begin{equation} 
\dot{h} \approx \sqrt{\frac{8}{\pi}} V_{\rm A0}
\left[
\ln \left(\frac{h}{\lambda_0}\right) - \frac{\pi}{2}
\right]^{\alpha}.
\label{eqn:CM_late_e}
\end{equation} 
The fitting parameters for CAlate\_e fit were
the critical height $\lambda_0$, 
Alfv\'en speed at the critical height $V_{\rm A0}$
and the exponent $\alpha$.
We also tested the original Eq.~\ref{eqn:CM_late} by
setting $\alpha$ equal to 0.5,
and the result will be called CAlate.
Because CA and CAlate\_e fits were based on non-linear differential equations,
we conducted the fittings by
numerically integrating the equations to 
produce a height--time relation,
which was then used to fit the data.
TI fit was implemented with the hyperbolic equation (Eq.~\ref{eqn:TI_early})
derived by \citet{KT2006PhRvL},
with $P_0$ and $P_1$ as the fitting parameters.
Poly fit was a polynomial of the following form:
\begin{equation}
h(t)= h_0 + v_0  (t-t_0) + a_0 (t-t_0)^{\alpha},
\label{eqn:poly}
\end{equation}
where $t_0$ corresponds to the time of the first data point.
The fitting parameters were
$h_0, v_0, a_0$ and $\alpha$.
The Poly fit not only can provide an empirical profile
for the later-stage kinematics,
but also may be used to distinguish the three models
because the proponents and/or modellers of TI, BO and CA
have all proposed 
such polynomial profile with 
different $\alpha$ for part of their respective CME kinematics
($\sim$3 for TI; \citet{Schrijver_etal2008ApJ},
$\sim$2 for BO (cf. Eq.~\ref{eqn:BO}), and
2.5 for CA (cf. Eq.~\ref{eqn:CM_early})).

As explained in Sec.~\ref{sec:mdl_CA} and \ref{sec:mdl_TI},
the profiles for CA, CAlate\_e and TI fits are 
the profiles for the early stage of the eruption.
Nevertheless, all four fitting models were applied to all the data sets.
The reason for fitting the higher coronal kinematics with 
CA, CAlate\_e and TI profiles
was that
we wanted to test the exclusiveness of these profiles to the initial stage.
If the kinematics of the later stages can also be well-fitted by these profiles,
the physical assumptions and conditions imposed to derive them may not be
sufficiently strict.
Our strategy of using these profiles to verify and discern different models
would thus be inappropriate and unreliable.
The data from different instruments were fitted separately because 
the full kinematics profile of a CME process is too complex to
be described by one single function.
In addition,
by fitting different data sets separately,
we avoided possible errors resulting from instrumental differences.

The fitting results of CME06 and CME07 are presented in
Fig.~\ref{fig:fits_cme061217} and \ref{fig:fits_cme071231}, respectively.
In the left panels of both figures,
we plotted only the results of the lower corona
to allow better visibility of the kinematics in this region.
The symbols along with the error bars in the figures
represent the results from the observational data,
and the continuous lines are the results from the fitting,
as denoted in the figures.

\section{Results} 
\subsection{2006 December 17 (CME06)} \label{sec:qual_cme06}
The initial part of this CME
matches the picture proposed by CA,
that is, the CME rose before the SXR emission began to increase.
The velocity profile
(cf. Fig.~\ref{fig:kin_cme061217}) also resembles the model profile
that corresponds to an
intermediate reconnection rate \citep[Fig.6 in][]{LinForbes2000JGR}.
However, a comparison of the kinematics and SXR emission profile
appears to indicate that there may be breakout reconnections 
before the formation of the current sheet.
This is explained as follows:
If the reconnections in the current sheet were indeed reflected in
the re-acceleration of the CME,
the rising in the SXR emission before this may appear to be
contradictory to the CA model.
Although the dissipation of the kinetic energy 
under an ideal MHD process can result in thermal emissions,
the energy is much lower than the energy of X-ray emissions.
An explanation we propose to explain this is that
there may be reconnections between the rising flux rope and 
the overlying magnetic fields
before the current sheet beneath the flux rope was formed.
While TI theory also proposed an ideal MHD process to initiate the instability,
most of the kinematic profiles of the CME leading edge from TI simulations
did not show a deceleration phase
\citep[see, e.g.,][]{KT2006PhRvL,TK2007AN328.743T,Schrijver_etal2008ApJ}.
However, we need further examinations to determine
whether TI may contribute to any part of this CME.

\begin{figure}
   \centering 
   \includegraphics[width=8cm]{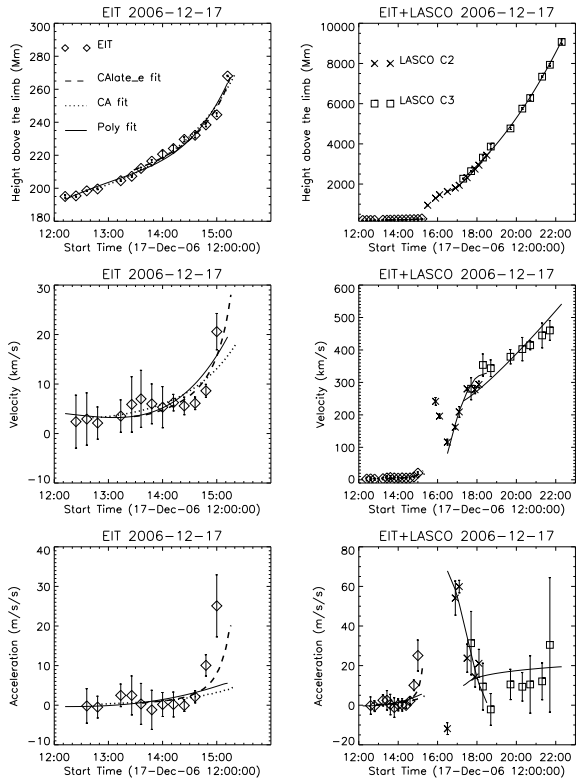}
   \caption{The fitting results of CME06.
            The best results of applying different fitting models
            (i.e., CA, CAlate\_e and poly) to 
            the height data of EIT and LASCO C2 and C3 are plotted
            in different line styles.
            Different symbols represent 
            the observational results of different instruments,
            as denoted in the figure.
            }
              \label{fig:fits_cme061217}%
    \end{figure}

The height--time curve of EIT can be fitted equally well
by CA, CAlate\_e and Poly expressions, but not TI.
Hence, only the fitting results of CA, CAlate\_e and Poly
were shown in Fig.~\ref{fig:fits_cme061217}.
The discrepancies among the three fittings only became visible
in the subsequently derived velocity and acceleration profiles.
Although all three fittings were consistent in the early part of the data,
only CAlate\_e re-produced the fast rising in the acceleration 
seen after $\sim$14:40~UT.
Since the CA-type expressions match the observational data,
we can use the CA theory to deduce that 
the eruptive flux rope seen in EIT may be free of current sheet
and that the current sheet may be formed shortly 
after the CME went out of the EIT FOV.
The critical height, $\lambda_0$, 
obtained from our CA fit was approximately 200~Mm,
the corresponding Alfv\'en speed, $V_{\rm A0}$, 
was between 130 and 250~km~s$^{-1}$,
and the initial speed was roughly 3~km~s$^{-1}$.
The observation showed that 
the foot-point separation of this CME was approximately 550 Mm
(cf. Fig.~\ref{fig:img_cme061217}).
Our fitting-determined critical height,
being roughly half of the foot-point separation,
is consistent with the CA theory (cf. Sec.~\ref{sec:mdl_CA}).
Assuming the plasma density 
$\rho \sim 
10^{-4}$kg~km$^{-3}$
and using $V_{\rm A0} \sim 200$~km~s$^{-1}$,
we can estimate the ambient magnetic field strength at the critical height
as follows:
\begin{eqnarray}
  B_{\lambda_0} &=& V_{\rm A0} \sqrt{4 \pi \rho} \nonumber \\
                &\approx& 7 \; \mbox{(G)},
\end{eqnarray}
which is a reasonable value at $\sim 200$~Mm above the limb.

The exponent $\alpha$ obtained from our Poly fit,
which gave a good match with the beginning part of the data,
was closer to 2 rather than 3,
which differed from the result by \citet{Schrijver_etal2008ApJ}.
Hence, the TI expressions produced by \citet{Schrijver_etal2008ApJ} 
and \citet{KT2006PhRvL} both failed to fit our data.
This Poly fit result, however,
cannot be used to reject either BO or CA because
$\alpha$ was mostly scattered between 2 and 2.5.

The best-fit profile of our best CAlate\_e fit had an exponent of 5.3. 
This leads to the following expression 
for the stage just before the formation of the current sheet:
\begin{equation}
\dot{h} \approx \sqrt{\frac{8}{\pi}} V_{\rm A0}
\left[
\ln \left(\frac{h}{\lambda_0}\right) - \frac{\pi}{2}
\right]^{5.3}.
\label{eqn:ftseq_EIT}
\end{equation}
$\lambda_0$ and $V_{\rm A0}$ given by the fitting
were $\sim 27$~Mm and 100~km~s$^{-1}$,
which appear to be inconsistent with the values from the CA fit.
However,
since the exponent in the expression is very different from 
the theoretically derived value 0.5 
(cf. Eq.~\ref{eqn:CM_late}),
$\lambda_0$ and $V_{\rm A0}$ from our CAlate\_e 
may no longer represent the critical height and Alfv\'en speed.
We found that setting the exponent to 0.5 failed to fit the data.
This indicates that while this CME may be driven by CA type instability,
the assumptions and approximations imposed to derive
Eq.~\ref{eqn:CM_late} were inadequate for this CME.

TI, CA and CAlate\_e expressions 
all failed to fit the LASCO data,
which is as expected 
because LASCO only observed the later stage of the CME event.
This is a promising indication that
the expressions for our quantitative examination are appropriate verification
tools.
Poly fit was the only one that rendered a close match to the height--time data.
However,
some differences between the fitting results and the data
become visible in the velocity and acceleration profiles.
Specifically, Poly fit was incapable of re-producing the dip in the C2 velocity.
Hence, Fig.~\ref{fig:fits_cme061217} shows 
the result of fitting only the later, linear part of C2.
In addition,
the acceleration profile derived from the Poly fit result 
failed to resemble the profile derived from the data.

The fitting results of EIT and LASCO
both revealed that
the discrepancies become larger 
in the higher order derivatives.
This suggests that
the derived properties (e.g., velocity and accelerations) 
are more discerning than the directly fitted property (e.g., height)
in verifying a model expression.

\subsection{2007 December 31 (CME07)} \label{sec:qual_cme07}
Although the initial eruption phase of this CME was occulted,
as we can infer from
the derived initial velocity and acceleration profile 
(cf. Fig.~\ref{fig:kin_cme071231}),
the peak acceleration in this phase is likely to coincide with
the maximum of the derivative of the SXR profile
\citep{Neupert1968ApJ,Temmer_etal2008ApJ}.
By a quick visual inspection of Fig.~\ref{fig:kin_cme071231},
we estimate that the peak occurred approximately between 00:45 and 00:50~UT,
which is also when a HXR emission was detected.

The kinematics profile of this CME and the SXR and HXR emission profiles
are in good agreement
with a BO scenario:
The SXR emissions and flaring activities
before the emergence of the loop
and the HXR signal detected on the top of the emerged loop
could be evidence of breakout reconnections.
The second HXR signal,
coinciding with the beginning of the re-acceleration phase,
indicated that
the reconnections
cut off the drag force from the current sheet beneath the CME loop
and accelerated the CME again.
However, 
because the very beginning rising stage of the CME
may have been obscured by the magnetic arcades,
the CME may have begun to rise earlier than 
the increase of the SXR emissions,
in which case, the timing would be in agreement with the CA theory.
Since the reconnection is not a necessary condition 
to either initiate or accelerate a CME in a TI picture,
we cannot use the X-ray emissions and flaring activities
to verify the TI model.
However, 
as we explained in Sec.~\ref{sec:qual_cme06},
most TI simulations of the CME leading-edge kinematics
did not show a drop of acceleration 
in the early phase of the eruption.
Based on the above discussion and reasoning,
we deduce that this CME was most likely driven by a BO mechanism.

   \begin{figure}
   \centering 
   \includegraphics[width=8cm]{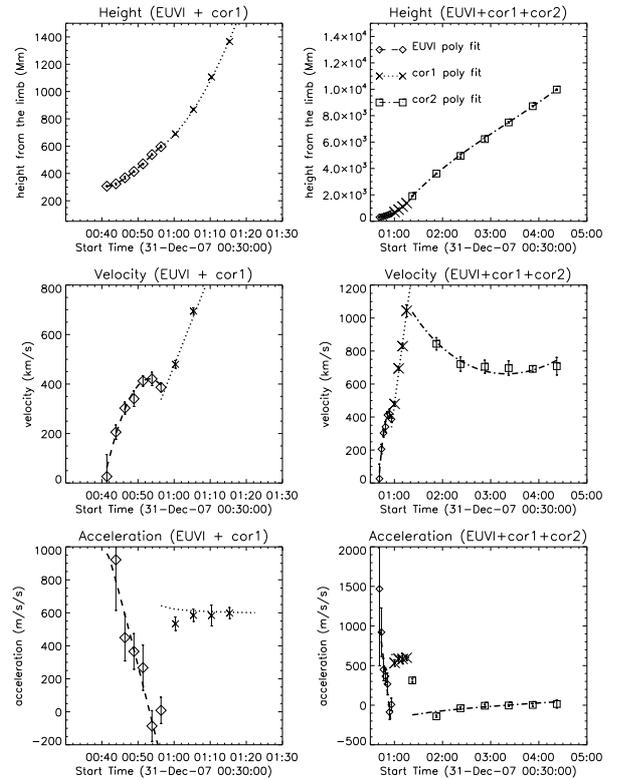}
   \caption{The fitting results of CME07.
            The fitting model is a polynomial of the form Eq.~\ref{eqn:poly}.
            Different symbols and continuous lines respectively represent 
            the observational and fitting results from
            different instruments,
            as denoted in the figure.
            }
              \label{fig:fits_cme071231}%
    \end{figure}

The kinematics of CME07 were well re-produced by the Poly fit, 
but not the other three models.
TI and CA-type expressions
failed to fit even the height data.
Hence, 
only the results of Poly fit were plotted in Fig.~\ref{fig:fits_cme071231}.
The unsuccessful fitting using 
the TI and CA-type expressions
is another indication that
the CME was driven by neither of the mechanisms 
and/or that the early stage was not detected,
which are both in agreement with
the results from our qualitative comparison.

As can be seen in Fig.~\ref{fig:fits_cme071231},
despite a few mismatching points in the velocity and acceleration, 
overall, 
the kinematics,
both the height and the derived velocity and acceleration,
of the observation and of the fitting
were in good agreement.
Hence, we can deduce that each stage of this CME can be approximated
as a polynomial of the form $h=h_0 + v_0*t  + a_0 * t^{\alpha}$
even though the kinematics of entire process is too complex to be
fitted by one single function.
The fitting results suggest that the order of polynomial (i.e., $\alpha$) for 
the height--time data from the three instruments
were $\sim 3$ for EUVI, $\sim 2$ for cor1 and $\sim 2.65$ for cor2.

\section{Conclusions}\label{sec:conclusion}

One of the main goals of this study was to investigate 
the possible driving mechanisms 
behind coronal mass ejections.
We have analyzed, both qualitatively and quantitatively, two CME events,
and compared them with three CME models for this purpose.

The first CME was observed on 2006 December 17 by EIT and LASCO,
and the associated X-ray emissions were recorded by GOES and RHESSI.
This CME lasted for over eight hours,
and the SXR emission exhibited two rising phases.
A comparison of the SXR emission profile
and the kinematics of the CME showed that
the CME was launched earlier than
the rising of the SXR emissions,
reached an acceleration peak during
the second rising phase of the SXR emission.
The CME was first decelerated after the eruption,
and then re-accelerated to another acceleration peak of $\sim$60~m~s$^{-2}$,
at which time the SXR emissions reached the maximum
and a flare was detected.
The CME propagated away at a small, constant acceleration at the end.

Comparing 
the above information
with the three CME models discussed in this paper,
CA, TI and BO,
we propose that 
this CME can be best described by a combination of CA and BO models.
That is, the CME was initiated by an ideal MHD process (CA),
as hinted by the absence of SXR emissions 
during the initial stage.
The rising and expanding of the flux rope 
against the ambient magnetic fields then 
led to reconnections
with these fields (BO).
As it continued to rise,
a current sheet was formed underneath,
which exerted a dragging force to slow down the rising motion,
and, lastly, the current sheet was cut off by reconnections
and the CME was re-accelerated.
Our quantitative examinations also revealed that
the CA-type expressions gave the best fit to the EIT data.
From the fitting results,
we deduced that the critical height at which the CME erupted
was approximately 200~Mm and 
the Alfv\'en speed at this height was around 130--250~km~s$^{-1}$,
which leads to a magnetic field of about 7 Gauss.
They are of reasonable order of magnitude for the low corona.
Based on the model results by  \citet{LinForbes2000JGR},
we can also infer 
from the occurrence of the deceleration of this CME
that the reconnection rate in the current sheet
may correspond to a Alfv\'en Mach number between 0.005 and 0.041.
%

The second CME was observed on 2007 December 31.
This CME was formed in a magnetic arcade system.
Our analysis indicated that the initial eruption stage might have been
obscured by the arcade system.
After this un-detected eruption,
the acceleration of the CME dropped first and 
then rose again to a peak value of $\sim$600~m~s$^{-2}$.
The CME remained at this peak acceleration until
its velocity reached a peak value of $\sim$1000~km~s$^{-1}$.
The CME then decelerated and eventually propagated away at 
a constant speed of $\sim$700~km~s$^{-1}$.
A comparison between the kinematics of the CME and 
the SXR and HXR emission information from 
GOES, MESSENGER SAX, and RHESSI
\citep[cf., Fig.~\ref{fig:kin_cme071231};][]{Raftery_etal2010apj,Krucker2009SPD,Krucker_etal2009ApJ}
showed 
that the SXR emission began to rise before the emergence of the CME
and peaked around the same time as the peak of the CME velocity,
and that
the HXR emission due to particle accelerations
occurred
at the time of the CME emergence and at the moment when the acceleration
of the CME began to rise again.

These observed features can be well described by a BO model as follows:
the SXR emission before the CME emerged from the magnetic arcades
and the high velocity and acceleration at the beginning of its emergence
can be explained as that
the breakout reconnections were happening 
as the CME loop was rising through and breaking the overlying arcade.
The reconnections accelerated the CME, and caused the X-ray emissions.
By the time the CME broke through the arcades,
a trailing current sheet may have been formed
to act as a drag force,
which was reflected in the decrease of the acceleration.
However, we note that
some overlying features, such as streamers,
could also slow down the CME.
The subsequent rising of the acceleration (after 01:00~UT)
and the detection of a HXR signal
indicated that this drag force was being reduced
by the reconnections in the current sheet.
After the current sheet was completely cut off,
the ambient coronal magnetic fields moved in to reform
and settle to a lower potential state.
The last stage of the CME indicated that it
was being slowed down by the solar wind and/or interstellar medium.
To be in agreement with a CA model,
which proposes an ideal MHD process to initiate the eruption,
a CME should have a beginning rising phase 
during which there was no X-ray emission.
This is inconsistent with the observations of this CME,
which detected flares and X-ray emissions 
from the very beginning of the eruption. 
However, since the initial rising stage of the CME was occulted,
we cannot completely rule out the possibility that
an ideal MHD process had taken place.
The missing of this ideal MHD process in our data
was also reflected in our unsuccessful fitting using CA-type expressions.

Both events exhibited a re-acceleration phase,
during which the maximum acceleration coincided with the peak SXR emission.
If higher SXR emissions indicate greater reconnection rates,
as explained in Sec.~\ref{sec:mdl_CA},
such coincidence can be explained as that
the faster reconnections,
which lead to faster reduction of the tethering force,
increased the acceleration of the CME.

In conclusion,
we have investigated two CME events.
Our results showed that 
the first event, CME06, can be best described by
a combination of the CA and BO models
while the other event, CME07, can be well explained by the BO model alone.
However,
it is also possible that CME07 may have been initiated by
an ideal MHD process.
We conclude that 
TI model is the least likely driving mechanism for either of the events
mainly because 
our observationally derived kinematic profiles
do not match the profiles and expressions from the TI model.
Specifically,
the decreasing of acceleration in the early stage of our CME events
is not seen in TI simulations to-date
\citep{KT2006PhRvL,TK2007AN328.743T,Schrijver_etal2008ApJ}.
Since our data is a two-dimensional projection of the actual motion,
we acknowledge the possibility that this phase be a projection effect. 
Hence, to improve the accuracy of our analysis,
we plan to utilize stereoscopic data to obtain an actual, 
three-dimensional kinematics of CMEs in our future work.
At last,
to improve the current CME models,
our study suggests that
the CA model include the reconnections/interactions 
between the CME and its ambient magnetic fields,
BO model consider the possibility of an ideal MHD process in the early stage,
and that TI model provide possible mechanisms for a drag force during
the eruptive phase.

\begin{acknowledgements}
We thank the referee for helpful comments that have improved the paper.
This work is supported by
an ESA/PRODEX grant administered by Enterprise Ireland.
CHL is also supported by National Center for Theoretical Sciences,
Physics Division, Taiwan.
\end{acknowledgements}

\bibliographystyle{aa.bst}
\bibliography{ref_cme}

\begin{thebibliography}{35}
\expandafter\ifx\csname natexlab\endcsname\relax\def\natexlab#1{#1}\fi

\bibitem[{{Alexander} {et~al.}(2002){Alexander}, {Metcalf}, \&
  {Nitta}}]{AMN_2002GeoRL}
{Alexander}, D., {Metcalf}, T.~R., \& {Nitta}, N.~V. 2002, \grl, 29, 41

\bibitem[{{Antiochos} {et~al.}(1999){Antiochos}, {DeVore}, \&
  {Klimchuk}}]{Breakout1999ApJ}
{Antiochos}, S.~K., {DeVore}, C.~R., \& {Klimchuk}, J.~A. 1999, \apj, 510, 485

\bibitem[{{Bateman}(1978)}]{Bateman1978book}
{Bateman}, G. 1978, {MHD instabilities} (Cambridge, Mass., MIT Press, 1978)

\bibitem[{{Bong} {et~al.}(2006){Bong}, {Moon}, {Cho}, {Kim}, {Park}, \&
  {Choe}}]{Bong_etal2006ApJ}
{Bong}, S.-C., {Moon}, Y.-J., {Cho}, K.-S., {et~al.} 2006, \apjl, 636, L169

\bibitem[{{Brueckner} {et~al.}(1995){Brueckner}, {Howard}, {Koomen},
  {Korendyke}, {Michels}, {Moses}, {Socker}, {Dere}, {Lamy}, {Llebaria},
  {Bout}, {Schwenn}, {Simnett}, {Bedford}, \& {Eyles}}]{LASCO1995SoPh}
{Brueckner}, G.~E., {Howard}, R.~A., {Koomen}, M.~J., {et~al.} 1995, \solphys,
  162, 357

\bibitem[{{Chen}(1989)}]{Chen_1989ApJ}
{Chen}, J. 1989, \apj, 338, 453

\bibitem[{{Cremades} \& {Bothmer}(2004)}]{CB2004A&A}
{Cremades}, H. \& {Bothmer}, V. 2004, \aap, 422, 307

\bibitem[{{Delaboudini{\`e}re} {et~al.}(1995){Delaboudini{\`e}re}, {Artzner},
  {Brunaud}, {Gabriel}, {Hochedez}, {Millier}, {Song}, {Au}, {Dere}, {Howard},
  {Kreplin}, {Michels}, {Moses}, {Defise}, {Jamar}, {Rochus}, {Chauvineau},
  {Marioge}, {Catura}, {Lemen}, {Shing}, {Stern}, {Gurman}, {Neupert},
  {Maucherat}, {Clette}, {Cugnon}, \& {van Dessel}}]{EIT1995SoPh}
{Delaboudini{\`e}re}, J., {Artzner}, G.~E., {Brunaud}, J., {et~al.} 1995,
  \solphys, 162, 291

\bibitem[{{DeVore} \& {Antiochos}(2008)}]{DVA2008ApJ}
{DeVore}, C.~R. \& {Antiochos}, S.~K. 2008, \apj, 680, 740

\bibitem[{{Forbes} \& {Isenberg}(1991)}]{FI1991ApJ}
{Forbes}, T.~G. \& {Isenberg}, P.~A. 1991, \apj, 373, 294

\bibitem[{{Forbes} \& {Priest}(1995)}]{FP1995ApJ}
{Forbes}, T.~G. \& {Priest}, E.~R. 1995, \apj, 446, 377

\bibitem[{{Gallagher} {et~al.}(2003){Gallagher}, {Lawrence}, \&
  {Dennis}}]{Gallagher_etal2003ApJ}
{Gallagher}, P.~T., {Lawrence}, G.~R., \& {Dennis}, B.~R. 2003, \apjl, 588, L53

\bibitem[{{Garcia}(1994)}]{GOES1994SoPh}
{Garcia}, H.~A. 1994, \solphys, 154, 275

\bibitem[{{Howard} {et~al.}(2008){Howard}, {Moses}, {Vourlidas}, {Newmark},
  {Socker}, {Plunkett}, {Korendyke}, {Cook}, {Hurley}, {Davila}, {Thompson},
  {St Cyr}, {Mentzell}, {Mehalick}, {Lemen}, {Wuelser}, {Duncan}, {Tarbell},
  {Wolfson}, {Moore}, {Harrison}, {Waltham}, {Lang}, {Davis}, {Eyles},
  {Mapson-Menard}, {Simnett}, {Halain}, {Defise}, {Mazy}, {Rochus}, {Mercier},
  {Ravet}, {Delmotte}, {Auchere}, {Delaboudiniere}, {Bothmer}, {Deutsch},
  {Wang}, {Rich}, {Cooper}, {Stephens}, {Maahs}, {Baugh}, {McMullin}, \&
  {Carter}}]{STEREO2008SSRv}
{Howard}, R.~A., {Moses}, J.~D., {Vourlidas}, A., {et~al.} 2008, Space Science
  Reviews, 136, 67

\bibitem[{{Kliem} \& {T{\"o}r{\"o}k}(2006)}]{KT2006PhRvL}
{Kliem}, B. \& {T{\"o}r{\"o}k}, T. 2006, Physical Review Letters, 96, 255002

\bibitem[{{Krall} {et~al.}(2001){Krall}, {Chen}, {Duffin}, {Howard}, \&
  {Thompson}}]{Krall_etal2001ApJ}
{Krall}, J., {Chen}, J., {Duffin}, R.~T., {Howard}, R.~A., \& {Thompson}, B.~J.
  2001, \apj, 562, 1045

\bibitem[{{Krucker} {et~al.}(2009{\natexlab{a}}){Krucker}, {Hudson}, {White},
  \& {Lin}}]{Krucker2009SPD}
{Krucker}, S., {Hudson}, H.~S., {White}, S.~M., \& {Lin}, R.~P.
  2009{\natexlab{a}}, in AAS/Solar Physics Division Meeting, Vol.~40, AAS/Solar
  Physics Division Meeting, 868

\bibitem[{{Krucker} {et~al.}(2009{\natexlab{b}}){Krucker}, {Hudson}, {White},
  {Masuda}, {Wuelser}, \& {Lin}}]{Krucker_etal2009ApJ}
{Krucker}, S., {Hudson}, H.~S., {White}, S.~M., {et~al.} 2009{\natexlab{b}},
  \apj, (submitted)

\bibitem[{{Lin} \& {Forbes}(2000)}]{LinForbes2000JGR}
{Lin}, J. \& {Forbes}, T.~G. 2000, \jgr, 105, 2375

\bibitem[{{Lin} {et~al.}(1998){Lin}, {Forbes}, {Isenberg}, \&
  {Demoulin}}]{Lin_etal1998ApJ}
{Lin}, J., {Forbes}, T.~G., {Isenberg}, P.~A., \& {Demoulin}, P. 1998, \apj,
  504, 1006

\bibitem[{{Lin} {et~al.}(2002){Lin}, {Dennis}, {Hurford}, {Smith}, {Zehnder},
  {Harvey}, {Curtis}, {Pankow}, {Turin}, {Bester}, {Csillaghy}, {Lewis},
  {Madden}, {van Beek}, {Appleby}, {Raudorf}, {McTiernan}, {Ramaty}, {Schmahl},
  {Schwartz}, {Krucker}, {Abiad}, {Quinn}, {Berg}, {Hashii}, {Sterling},
  {Jackson}, {Pratt}, {Campbell}, {Malone}, {Landis}, {Barrington-Leigh},
  {Slassi-Sennou}, {Cork}, {Clark}, {Amato}, {Orwig}, {Boyle}, {Banks},
  {Shirey}, {Tolbert}, {Zarro}, {Snow}, {Thomsen}, {Henneck}, {McHedlishvili},
  {Ming}, {Fivian}, {Jordan}, {Wanner}, {Crubb}, {Preble}, {Matranga}, {Benz},
  {Hudson}, {Canfield}, {Holman}, {Crannell}, {Kosugi}, {Emslie}, {Vilmer},
  {Brown}, {Johns-Krull}, {Aschwanden}, {Metcalf}, \&
  {Conway}}]{RHESSI2002SoPh}
{Lin}, R.~P., {Dennis}, B.~R., {Hurford}, G.~J., {et~al.} 2002, \solphys, 210,
  3

\bibitem[{{Lynch} {et~al.}(2008){Lynch}, {Antiochos}, {DeVore}, {Luhmann}, \&
  {Zurbuchen}}]{Lynch_etal2008ApJ}
{Lynch}, B.~J., {Antiochos}, S.~K., {DeVore}, C.~R., {Luhmann}, J.~G., \&
  {Zurbuchen}, T.~H. 2008, \apj, 683, 1192

\bibitem[{{Lynch} {et~al.}(2004){Lynch}, {Antiochos}, {MacNeice}, {Zurbuchen},
  \& {Fisk}}]{Lynch_etal2004ApJ}
{Lynch}, B.~J., {Antiochos}, S.~K., {MacNeice}, P.~J., {Zurbuchen}, T.~H., \&
  {Fisk}, L.~A. 2004, \apj, 617, 589

\bibitem[{{Manoharan} \& {Kundu}(2003)}]{MK2003ApJ}
{Manoharan}, P.~K. \& {Kundu}, M.~R. 2003, \apj, 592, 597

\bibitem[{{Neupert}(1968)}]{Neupert1968ApJ}
{Neupert}, W.~M. 1968, \apjl, 153, L59

\bibitem[{{Priest} \& {Forbes}(2000)}]{PFbook2000}
{Priest}, E. \& {Forbes}, T. 2000, {Magnetic Reconnection: MHD Theory and
  Applications} (Magnetic Reconnection, by Eric Priest and Terry Forbes,
  pp.~359.~ISBN 0521481791.~Cambridge, UK: Cambridge University Press, June
  2000.), 359--393

\bibitem[{{Priest} \& {Forbes}(2002)}]{PF_2002A&ARv}
{Priest}, E.~R. \& {Forbes}, T.~G. 2002, \aapr, 10, 313

\bibitem[{{Raftery} {et~al.}(2010){Raftery}, {Gallagher}, {McAteer}, {Lin}, \&
  {Delahunt}}]{Raftery_etal2010apj}
{Raftery}, L.~C., {Gallagher}, R.~T., {McAteer}, R.~T.~J., {Lin}, C.-H., \&
  {Delahunt}, G. 2010, \apj, (submitted)

\bibitem[{{Reeves} \& {Forbes}(2005)}]{ReevesForbes2005ApJ}
{Reeves}, K.~K. \& {Forbes}, T.~G. 2005, \apj, 630, 1133

\bibitem[{{Schrijver} {et~al.}(2008){Schrijver}, {Elmore}, {Kliem},
  {T{\"o}r{\"o}k}, \& {Title}}]{Schrijver_etal2008ApJ}
{Schrijver}, C.~J., {Elmore}, C., {Kliem}, B., {T{\"o}r{\"o}k}, T., \& {Title},
  A.~M. 2008, \apj, 674, 586

\bibitem[{{Sheeley} {et~al.}(1999){Sheeley}, {Walters}, {Wang}, \&
  {Howard}}]{Sheeley_etal1999}
{Sheeley}, N.~R., {Walters}, J.~H., {Wang}, Y.-M., \& {Howard}, R.~A. 1999,
  \jgr, 104, 24739

\bibitem[{{Temmer} {et~al.}(2008){Temmer}, {Veronig}, {Vr{\v s}nak},
  {Ryb{\'a}k}, {G{\"o}m{\"o}ry}, {Stoiser}, \& {Mari{\v
  c}i{\'c}}}]{Temmer_etal2008ApJ}
{Temmer}, M., {Veronig}, A.~M., {Vr{\v s}nak}, B., {et~al.} 2008, \apjl, 673,
  L95

\bibitem[{{T{\"o}r{\"o}k} \& {Kliem}(2007)}]{TK2007AN328.743T}
{T{\"o}r{\"o}k}, T. \& {Kliem}, B. 2007, Astronomische Nachrichten, 328, 743

\bibitem[{{van Tend} \& {Kuperus}(1978)}]{VanTend_Kuperus1978SoPh}
{van Tend}, W. \& {Kuperus}, M. 1978, \solphys, 59, 115

\bibitem[{{Zhang} {et~al.}(2004){Zhang}, {Dere}, {Howard}, \&
  {Vourlidas}}]{Zhang_etal2004ApJ}
{Zhang}, J., {Dere}, K.~P., {Howard}, R.~A., \& {Vourlidas}, A. 2004, \apj,
  604, 420

\end{thebibliography}

\end{document}